\documentclass[prd,amssymb,twocolumn,nofootinbib,10pt]{revtex4-1}
\pdfoutput=1
\usepackage{amsmath,amsfonts,amssymb}
\usepackage{color,verbatim,graphicx,float}
\usepackage{bbold}

\begin{document}

\small{\textsc{Version accepted for publication in Phys. Rev. D.}}

\title{Minisuperspace model of compact phase space gravity}

\author{Danilo Artigas$^{*,\ddagger}$}
\author{Jakub Mielczarek${}^{*,\dagger}$}
\author{Carlo Rovelli${}^{*}$}
\affiliation{
${}^*$Aix Marseille Universit\'e, Universit\'e de Toulon, CNRS, CPT, 13288 Marseille, France \\ 
${}^\ddagger$Magist\`{e}re de Physique Fondamentale, Univ. Paris-Sud, Univ.  Paris-Saclay, 91405 Orsay, France\\
${}^\dagger$Institute of Physics, Jagiellonian University, ul.\ {\L}ojasiewicza 11, 30-348 Krak\'{o}w, Poland
}

\begin{abstract}
The kinematical phase space of classical gravitational field is flat (affine) and unbounded. 
Because of this, field variables may tend to infinity leading to appearance of singularities, 
which plague Einstein's theory of gravity. The purpose of this article is to study the 
idea of generalizing the theory of gravity by compactification of the phase space. We investigate 
the procedure of compactification of the phase space on a minisuperspace
gravitational model with two-dimensional phase space. In the affine limit, the model reduces
to the flat de Sitter cosmology. The phase space is generalized to the spherical case, and 
the case of loop quantum cosmology is recovered in the cylindrical phase space limit. Analysis 
of the dynamics reveals that the compactness of the phase space leads to both UV and 
IR effects. In particular, the phase of recollapse appears, preventing the Universe from 
expanding to infinite volume. Furthermore, the quantum version of the model is investigated and 
the quantum constraint is solved. As an example, we analyze the case with the spin quantum 
number $s=2$, for which we determine transition amplitude between initial and final state of 
the classical trajectory. The probability of the transition is peaked at $\Lambda=0$. 
\end{abstract}  

\maketitle

\section{Introduction}

Compact phase spaces emerge in the semiclassical description of a quantum system with 
finite dimensional Hilbert spaces. An important property of compactness is that values of 
phase space variables are bounded. As a consequence, physical quantities such as 
energy density may be constrained from above, resolving the problem of divergences 
appearing in the case of affine phase spaces. Compactification of phase spaces may, therefore, 
serve as a way to impose the principle of finiteness, introduced by Max Born and 
Leopold Infeld \cite{Born:1934fy}.  

Following this reasoning, in Ref. \cite{Mielczarek:2016rax} a research program of nonlinear 
field space theory (NFST) has been initiated with the goal of generalizing the known types
of physical fields to the case of compact phase spaces. In the original article \cite{Mielczarek:2016rax} 
the procedure of compactification has been investigated at the level of the Fourier space 
representation of a scalar field theory. For the standard scalar field, each Fourier mode is 
associated with a two-dimensional $\mathbb{R}^2$ phase space, which in the NFST has been
considered as a local approximation to the spherical phase space $\mathbb{S}^2$. The 
procedure has been thereafter applied for the field defined in the position space in Ref. 
\cite{Mielczarek:2016xql}. It has been shown that, thanks to the fact that the spherical 
phase space is a phase space of angular momentum (spin), a new possibility of relating 
spin systems with field theories emerges. In particular, it has been shown in Ref. \cite{Mielczarek:2016xql}
that small excitations of the continuous version of the Heisenberg model are described 
by nonrelativistic scalar field theory, if the large spin limit ($S\rightarrow \infty$) is 
considered.  Furthermore, the scalar field theory recovered satisfies the so-called 
Born reciprocity symmetry between generalized positions and conjugate 
momenta \cite{Born49}. Worth mentioning is that, while the Born reciprocity
has mostly been forgotten for years because of its incompatibility with relativistic 
symmetries, it has recently reemerged in the context of string theory and quantum 
gravity \cite{Freidel:2013zga,Freidel:2014qna}.

The next step of the NFST program was to show that the construction 
can be generalized to the case of relativistic Klein-Gordon scalar field theory 
\cite{Bilski:2017gic}. It was demonstrated that such theory is recovered in the large 
spin limit of the XX Heisenberg model (XXZ Heisenberg model in the limit of the 
vanishing anisotropy parameter $\Delta \rightarrow 0$). Possible consequences of 
the compact phase space scalar field theory have been investigated in the cosmological
context, by applying the compactness to the inflationary scalar field 
\cite{Mielczarek:2017ny}. It has been shown that the compactness of the inflaton field 
may have implications on amplitudes of cosmological primordial inhomogeneities as 
well as leading to the phase of cosmological recollapse \cite{Mielczarek:2017ny}.   

While the case of scalar field theory is a good testing ground for the procedure of 
compactification of the phase spaces, the ambition of the NFST is to ultimately apply 
the method to the gravitational interactions. The expectation is that compact 
phase space extension of general relativity (GR) may resolve the problem of singularities 
(simply by restricting field variables to take finite values) and pave a way to a finite 
Hilbert space quantum version of the theory of gravity. A possibility of such an approach 
to quantum theory of gravity has already been discussed in the context of loop quantum 
gravity (LQG) in Ref. \cite{Rovelli:2015fwa}. In the current formulation, LQG is a theory 
with $su(2)\times SU(2)$ phase space per link of the spin network. Part of the phase space, 
associated with the $SU(2)$ holonomies is already compact. However, the remaining 
contribution is affine and is described by elements of the $su(2)$ algebra. The idea 
pushed forward in Ref. \cite{Rovelli:2015fwa} was that generalization of the theory to 
the compact phase space $SU(2)\times SU(2)$ may resolve certain problems 
(e.g. IR divergences) present in the current formulation.  Furthermore, nontrivial phase 
spaces are considered in the relative locality approach to quantum gravity 
\cite{AmelinoCamelia:2011bm}. However, at the current stage of development, the  
phase space of particles rather than field (including gravitational field) is considered 
in this approach. 
 
The purpose of this article is to make a step towards construction of the compact 
phase space version of GR by studying the procedure of compactification of the 
gravitational degrees of freedom in a minisuperspace model. More specifically,
our objective is to introduce phase space compactness to the flat 
Friedmann-Robertson-Walker (FRW) cosmological model with positive cosmological 
constant $\Lambda$. There is of course a freedom of choices of possible compact phase 
space extensions of initial affine phase space theory. In our studies, we explore 
the spherical $\mathbb{S}^2$ phase space, which will allow us to build relations 
with the spin physics. However, in general, other possibilities, such as toroidal 
phase space $\mathbb{S}\times \mathbb{S}$, can also be considered. 

We study the classical and the quantum theory for which exact analytical solutions 
are found. We show that loop quantum cosmology is recovered in a certain limit. 
We demonstrate how transition amplitude can be explicitly computed in the quantum 
theory using the projector onto solutions of the quantum Hamiltonian constraint and 
coherent boundary states in the kinematical Hilbert space.    

The organization of the article is as follows. In Sec. \ref{SecdeSittermodel} we introduce 
the standard de Sitter model and notation used through this article is established. 
Then in Sec. \ref{SphericalPS} the spherical phase space is introduced and the 
affine large spin limit at the kinematical level is discussed.  Based on this, in Sec. 
\ref{SecCompactFRW} the standard FRW dynamics with positive cosmological 
constant is generalized to the case with compact phase space. The Hamiltonian 
constraint we obtain is expressed in terms of the spin variable, which parametrizes 
the phase space. In Sec. \ref{SecPolymerLimit} we show that the theory 
reduces to loop quantum cosmology if the phase space is elongated in the direction 
of one of the canonical variables. This yields a cylindrical phase space. In 
Sec. \ref{CosmologicalEvolution} we derive analytical solutions of the full model. 
The quantum analysis of the model is performed in Sec. \ref{QuantumConsiderations},
where the quantum constraint is explicitly solved and exemplary transition amplitudes 
associated with end points of the classical trajectory are calculated. The results 
are summarized in Sec. \ref{SecSummary}.

\section{de Sitter model}
\label{SecdeSittermodel}

The phase space is a symplectic manifold equipped with closed 2-form $\omega$.
In the case of the FRW cosmology, $\omega$ can be written in the Darboux form
\begin{equation}
\omega = dp \wedge dq, \label{2formFRW} 
\end{equation}
where $q$ is the generalized coordinate and $p$ is the canonically conjugated 
momentum. $q$ is a volume element, related to the scale factor $a$ and fiducial 
volume $V_0$ as follows: $|q|:=V_0 a^3$. Keeping in mind different possible triad 
orientations, we allow both positive and negative real values of $q$. As a 
consequence, the phase space for the system is $\mathbb{R}^2$. 

In terms of the $q$ and $p$ variables, the Hamiltonian for the flat FRW cosmology with 
cosmological constant $\Lambda$ takes the form \cite{Mielczarek:2017ny}
\begin{equation}
H_{\text{GR}}=Nq\left(- \frac{3}{4}\kappa p^2+ \frac{\Lambda}{\kappa}\right)
\label{FRWHamiltonian}
\end{equation}
where $\kappa := 8\pi G = 8 \pi l^2_{\text{Pl}}$ and $N$ is the lapse function.  
$l_{\text{Pl}} \approx 1.62 \cdot 10^{-35}$ is the Planck length. 

By inverting the symplectic form (\ref{2formFRW}) the Poisson bracket 
 \begin{equation}
\{f,g\} =(\omega^{-1})^{ij}\partial_i f \partial_j g  
= \frac{\partial f}{\partial q} \frac{\partial g}{\partial p}-\frac{\partial f}{\partial p} \frac{\partial g}{\partial q}. 
\label{Poisson1}
\end{equation} 
can be introduced, where $f$ and $g$ are phase space functions. 
The Poisson bracket allows us to introduce the Hamilton equation $\dot{f} = 
\{f, H_{\text{GR}}\}$.  In the cosmological context it is useful to introduce 
the Hubble factor $H$, which quantifies the rate of cosmological expansion.
\begin{equation}
H := \frac{1}{3} \frac{\dot{q}}{q}, 
\label{HubbleFactor}
\end{equation}
where the time derivative of $q$ is defined in terms of the Hamilton equation 
$\dot{q} = \{q, H_{\text{GR}}\} = - \frac{3}{2} N \kappa qp$. Fixing, 
from now on, the gauge $N=1$, we can express the Hubble factor as follows:
\begin{equation}
H=-\frac{1}{2}\kappa p. 
\label{HPeq}
\end{equation}
Plugging this relation into the scalar constraint, we have 
\begin{equation}
0=\frac{\partial H_{\text{GR}}}{\partial N} = q\left(- \frac{3}{4}\kappa p^2+ \frac{\Lambda}{\kappa}\right),
\end{equation} 
and the Friedmann equation in the well-known form 
\begin{equation}
H^2= \frac{\Lambda}{3} 
\label{FriedmannFlat}
\end{equation}
is recovered, with exponential solutions $q(t)=C e^{\pm \sqrt{3 \Lambda}t}$ for $N=1$.

\section{Spherical phase space}
\label{SphericalPS}

For the spherical phase space $\mathbb{S}^2$ the natural candidate 
for the symplectic 2-form is the area form,
\begin{equation}
\omega = S \sin\theta\, d\phi \wedge d\theta,
\label{sphere2form1}
\end{equation}
where $\theta$ and $\phi$ are spherical angles and $S$ has been introduced 
for dimensional reasons.  The volume (area) of the phase space is now finite 
and equal to 
\begin{equation}
A = \int_{\Omega} \omega = 4 \pi S.  
\end{equation}
The affine limit corresponds to $S \rightarrow \infty$. As showed in Ref. \cite{Mielczarek:2016xql}, 
it is convenient to perform a change of coordinates in the form:
\begin{eqnarray}
\phi &=& \frac{p}{R_1}  \in (-\pi, \pi], \\
\theta &=& \frac{\pi}{2}+\frac{q}{R_2}  \in (0,\pi), 
\end{eqnarray}
such that the 2-form (\ref{sphere2form1}) rewrites as 
\begin{equation}
\omega = \cos\left( \frac{q}{R_2}\right) dp \wedge dq, \label{S2form4}
\end{equation}
where we have set that $R_1 R_2 = S$. This guarantees that in the 
large $S$ limit ($R_{1,2}\rightarrow \infty$), the symplectic form (\ref{S2form4}) simplifies
to the $\mathbb{R}^2$ case with  symplectic form (\ref{2formFRW}). The form 
(\ref{S2form4}) differs from the one introduced in Ref. \cite{Mielczarek:2016xql} by the 
change of variables $(q \rightarrow p, p\rightarrow -q)$, which does not change 
physics but the convention used here allows us to make direct connection with polymerization 
of momentum $p$. Based on the symplectic form (\ref{S2form4}), the 
Poisson bracket becomes  
\begin{equation}
\{f,g\} = \frac{1}{\cos(q/R_2)}\left( \frac{\partial f}{\partial q} \frac{\partial g}{\partial p}
-\frac{\partial f}{\partial p} \frac{\partial g}{\partial q}\right). 
\label{PoissonS2}
\end{equation} 
The difference with the $\mathbb{R}^2$ case (\ref{Poisson1}) is the presence 
of the factor $1/\cos(q/R_2)$.  Both the $q$ and $p$ variables are only locally 
well defined on the sphere, but it is justified to introduce globally defined 
functions to study the dynamics also away from the origin of the coordinate  
system $(q,p)=(0,0)$. A choice that guarantees that the new variables
are globally defined and the algebra of the variables takes a simple form 
is associated with the parametrization of the sphere in a Cartesian coordinate 
system. Namely, we introduce a vector $\vec{S}=(S_x,S_y,S_z)$, with components 
expressed in terms of the $q$ and $p$ variables as follows:
\begin{align}
S_x&=S\cos\left(\frac{p}{R_1}\right)\cos\left(\frac{q}{R_2}\right),\\
S_y&=S\sin\left(\frac{p}{R_1}\right)\cos\left(\frac{q}{R_2}\right),\\
S_z&=-S\sin\left(\frac{q}{R_2}\right).
\label{S_z}
\end{align}
Differentiating the components with respect to the $q$ and $p$  variables we get
\begin{equation}
\frac{\partial S_i}{\partial q}=
\left\{
  \begin{array}{l c l}
\partial S_x/\partial q=-R_1\cos(p/R_1)\sin(q/R_2)\\
\partial S_y/\partial q=-R_1\sin(p/R_1)\sin(q/R_2)\\
\partial S_z/\partial q=-R_1\cos(q/R_2)
  \end{array}
   \right. ,
\end{equation}
and 
\begin{equation}
\frac{\partial S_i}{\partial p}=
\left \{
  \begin{array}{l c l}
\partial S_x/\partial p=-S_y/R_1\\
\partial S_y/\partial p=S_x/R_1\\
\partial S_z/\partial p=0
  \end{array}
   \right. \ ,
\end{equation}
when applied to the Poisson bracket (\ref{PoissonS2}), we find that the $S_i$ components 
satisfy the $\mathfrak{so(3)}$ algebra
\begin{equation}
\{S_i,S_j\}=\epsilon_{ijk}S_k.
\end{equation}
The vector $\vec{S}=(S_x,S_y,S_z)$ is therefore a vector of angular momentum (spin)
with magnitude equal to $S$. The affine limit $R_{1,2}\rightarrow \infty$ is, therefore, a 
large spin limit.  

\section{Compact phase space FRW model}
\label{SecCompactFRW}

In the previous section we have shown that, at the kinematical level, compactification 
of the affine phase space $\mathbb{R}^2$ of the FRW model to the case of $\mathbb{S}^2$
can be performed replacing the symplectic form (\ref{2formFRW}) with (\ref{S2form4}).
The second step is to introduce the compactness at the level of the dynamics, by suitable 
modification of the minisuperspace Hamiltonian (\ref{FRWHamiltonian}). Since the original 
$q$ and $p$ phase space variables are not globally defined on the sphere, we have to 
replace them with the $S_i$ variables introduced in the previous section. The consistency 
requirement is that in the large spin limit ($R_{1,2}\rightarrow \infty$) the classical FRW 
Hamiltonian (\ref{FRWHamiltonian}) would be recovered. We take here the additional requirement 
that in the $R_2\rightarrow \infty$ limit the case of the polymer quantization is recovered. 

The simplest way to satisfy the above conditions is to perform the following replacements: 
\begin{align}
p &\rightarrow p_S := \frac{S_y}{R_2} = R_1 \sin\left(\frac{p}{R_1}\right)\cos\left(\frac{q}{R_2}\right), \\
q &\rightarrow q_S:= -\frac{S_z}{R_1}=  R_2 \sin\left(\frac{q}{R_2}\right),  \label{defqS}
\end{align}
where the new variables are defined such that $\lim_{R_{1,2} \rightarrow \infty} 
p_S = p$ and $\lim_{R_{1,2}\rightarrow \infty} q_S = q$.  Applying the above 
replacements in Eq.  (\ref{FRWHamiltonian}) a new Hamiltonian defined on spherical 
phase space can be introduced,
\begin{equation}
H_{S}=N\frac{S_z}{R_1}\left[\frac{3}{4}\kappa\frac{S_y^2}{R_2^2}-\frac{\Lambda}{\kappa}\right],
\label{HamiltonianSphere}
\end{equation}
such that $H_S \rightarrow H_{\text{GR}}$ in the $R_{1,2}\rightarrow\infty$ limit.

The Hamiltonian (\ref{HamiltonianSphere}) and the Poisson bracket (\ref{PoissonS2})
yield the equations of motion:
\begin{align}
\dot{S_x}&=\frac{3}{4}\frac{N\kappa}{R_1 R_2^2}S_y\left(2S_z^2-S_y^2\right)
+\frac{N\Lambda}{R_1\kappa}S_y, \label{dotSx} \\
\dot{S_y}&=\frac{3}{4}\frac{N\kappa}{R_1 R_2^2}S_xS_y^2
-\frac{N\Lambda}{R_1\kappa}S_x, \label{dotSy} \\
\dot{S_z}&=-\frac{3}{2}\frac{N\kappa}{R_1 R_2^2} S_x S_y S_z. \label{dotSz}
\end{align}
With the use of the set of equations above, one can check that 
\begin{equation}
\frac{dS^2}{dt}=2(S_x\dot{S_x}+S_y\dot{S_y}+S_z\dot{S_z})=0. 
\end{equation}

Let us now derive the Friedmann equation. For this purpose, using the Poisson 
bracket (\ref{PoissonS2}), we calculate  
\begin{equation}
\dot{q}=  \frac{1}{\cos(q/R_2)} \frac{\partial H_S}{\partial p}
=\frac{3}{2}\frac{N\kappa}{S^2}\frac{S_xS_yS_z}{\cos(q/R_2)},
\end{equation}
so that (fixing as above the gauge $N=1$) the Hubble factor (\ref{HubbleFactor}) 
takes the following form 
\begin{equation}
H=\frac{1}{2}\frac{\kappa}{S^2q}\frac{S_xS_yS_z}{\cos(q/R_2)},
\label{HubbleSphere}
\end{equation}
or equivalently 
\begin{equation}
H=-\frac{S\kappa}{2q}\sin\left(\frac{p}{R_1}\right)\cos\left(\frac{p}{R_1}\right)
\sin\left(\frac{q}{R_2}\right)\cos\left(\frac{q}{R_2}\right), 
\end{equation}
such that in the $R_{1,2}\rightarrow \infty$ limit we recover 
$H=-\frac{1}{2}\kappa p$, as expected [see Eq. (\ref{HPeq})]. 

The $\frac{\partial H_S}{\partial N}=0$ condition implies that the scalar constraint 
takes the form
\begin{equation} 
\Lambda=\frac{3}{4}\kappa^2 \frac{S_y^2}{R_2^2},
\label{Lambda}
\end{equation}
which reduces to
\begin{equation} 
\sin^2\left(\frac{p}{R_1}\right)\cos^2\left(\frac{q}{R_2}\right) = \frac{4}{3} \frac{\Lambda}{\kappa^2R_1^2}.
\end{equation}
Therefore, the Friedmann equation takes the form
\begin{equation}
H^2 = \frac{\Lambda}{3}\left(\frac{\sin(q/R_2)}{q/R_2} \right)^2 
\left[\frac{\cos^2\left(q/R_2\right)-\delta}{\cos^2(q/R_2)}\right],
\label{FriedmannSphere}
\end{equation}
in the $N=1$ gauge. The real solutions to the equation can be obtained if the condition 
\begin{equation}
\cos^2\left(q/R_2\right) \geq \delta
\end{equation}
is satisfied, where for convenience we have defined 
\begin{equation}
\delta := \frac{4}{3} \frac{\Lambda}{R_1^2\kappa^2}.
\end{equation}

\section{Polymer limit}
\label{SecPolymerLimit}

\begin{figure}[ht!]
\centering
\includegraphics[width=7cm, angle = 0]{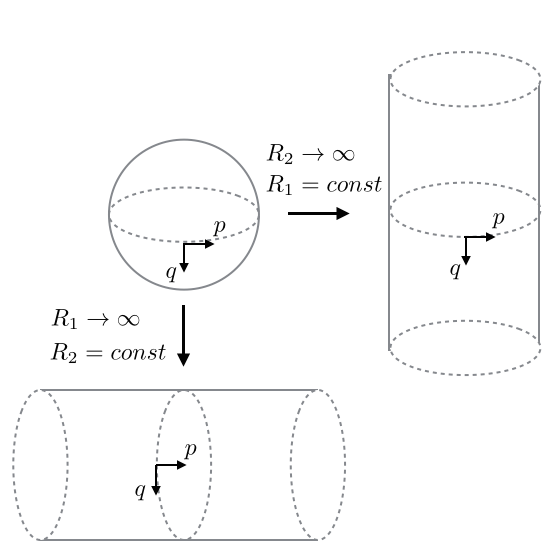}
\caption{Momentum and position polymerization (cylindrical) limits of the spherical phase space.}
\label{fig:Polymer}
\end{figure}

While the spherical case imposes restrictions on the values of $p$ and $q$, we now study 
the limits where the constraints on the phase space variables are released. This is equivalent 
to elongating the spherical phase space into a cylindrical shaped phase space, by taking either the 
$R_2\rightarrow\infty$ or $R_1\rightarrow\infty$ limit (see Fig. \ref{fig:Polymer}). The cylindrical 
phase space obtained corresponds to the so-called polymerization \cite{Ashtekar:2002sn}, 
playing a crucial role in loop quantum cosmology (LQC) \cite{Bojowald:2008zzb,Ashtekar:2011ni}. 
A preliminary analysis of the relation between spherical phase space and the polymerization 
of momentum at the kinematical has been performed in Ref. \cite{Bilski:2017gic}. Here, we 
consider the polymer limit in both canonical directions and explore consequences on dynamics.     

\subsection{Momentum polymerization}

We first consider the $R_2\rightarrow\infty$ limit, so that the symplectic form (\ref{S2form4}) reduces 
to the Darboux form (\ref{2formFRW}) and the Poisson bracket (\ref{PoissonS2}) reduces to the 
affine case (\ref{Poisson1}). Under this limit, the spin components reduce to
\begin{align}
S_x&=R_1R_2\cos\left(\frac{p}{R_1}\right),\\
S_y&=R_1R_2\sin\left(\frac{p}{R_1}\right),\\
S_z&=-R_1q,
\end{align}
together with $R_2\rightarrow\infty$. Based on the $\mathfrak{so(3)}$ algebra, this one reduces to
\begin{align} 
\left\{ \sin\left(\frac{p}{R_1}\right),\cos\left(\frac{p}{R_1}\right) \right\} &=0, \\
\left\{q, R_1\sin\left(\frac{p}{R_1}\right) \right\} &=\cos\left(\frac{p}{R_1}\right), \label{S1algebra2} \\
\left\{q,\cos\left(\frac{p}{R_1}\right) \right\} &=- \frac{1}{R_1}\sin\left(\frac{p}{R_1}\right),
\end{align}
which is the cylindrical algebra on the ${\mathbb{S}^1}\times\mathbb{R}$ phase space 
\cite{Bojowald:2011jd}. If one takes the  $R_1\rightarrow\infty$ limit the only nontrivial 
contribution to the algebra that remains is (\ref{S1algebra2}), which simplifies to 
$\left\{q, p \right\}=1$ as expected. 

The Hamiltonian of the spherical phase space (\ref{HamiltonianSphere}) reduces now to
\begin{equation}
H_{\text{C}}=Nq\left[-\frac{3}{4}R_1^2\kappa \sin^2\left(\frac{p}{R_1}\right)+\frac{\Lambda}{\kappa}\right],
\end{equation}
where the $C$ index denotes that we now deal with cylindrical phase space. 
We can simplify the Hamiltonian $H_{\text{C}}$ to
\begin{equation}
H_{\text{C}}=Nq\left[-\frac{3\kappa}{4}\frac{\sin^2\left(\lambda p\right)}{\lambda^2}
+\frac{\Lambda}{\kappa}\right], 
\label{HamiltonianR2Lim}
\end{equation}
where we have introduced the scale of polymerization $\lambda:= \frac{1}{R_1}$, 
so that in the limit $R_1\rightarrow \infty$ ($\lambda \rightarrow 0$) the FRW 
Hamiltonian $H_{\text{GR}}$ is recovered.

The expression (\ref{HamiltonianR2Lim}) is equivalent to the Hamiltonian of LQC, where we have
the following polymerization of the momentum:
\begin{equation}
p \rightarrow p_{\lambda} := \frac{\sin\left(\lambda p\right)}{\lambda},
\end{equation}
such that $\lim_{\lambda\rightarrow 0} p_{\lambda} = p$.

In the $R_2\rightarrow \infty$ limit, the Friedmann equation (\ref{FriedmannSphere}) simplifies to
\begin{align}
H^2 &=\frac{\Lambda}{3}\left(1- \delta \right)
\label{FreidmannPolymer1}
\end{align}
for $N=1$, and a real solution to the equation can be found only if $\delta \in [0,1]$. 
Hence, the dynamics imposes bounds on the possible values of cosmological 
constant. This effect has already been noticed before in the polymerized cosmology 
\cite{Mielczarek:2008zv,Mielczarek:2010rq,Mielczarek:2010wu}.  

Using the expression for the energy density of the cosmological constant 
$\rho_{\Lambda}:= \frac{\Lambda}{\kappa}$, the Friedmann equation 
(\ref{FreidmannPolymer1}) can be rewritten as
\begin{align}
H^2 &=\frac{\kappa}{3}\rho_{\Lambda} \left(1-\frac{\rho_{\Lambda}}{\rho_{c}}\right), 
\label{FreidmannPolymer2}
\end{align}
where the $\rho_c$ is the critical energy density considered in LQC,
\begin{equation}
\rho_c := \frac{3}{4}\frac{\kappa}{\lambda^2}.
\end{equation}
The leading contribution in Eq. (\ref{FreidmannPolymer2}) matches with the Friedmann equation for a flat 
phase space (\ref{FriedmannFlat}), whereas the second contribution implies correction to the cosmological 
evolution due to the cylindrical phase space. Equation (\ref{FreidmannPolymer2}) has a solution in the exponential 
form 
\begin{equation}
q(t)=C e^{\pm \sqrt{3 \Lambda_{\text{eff}}}t}
\label{deSitterEff}
\end{equation}
where we have introduced an effective cosmological constant  
$\Lambda_{\text{eff}} := \Lambda\left(1-\delta \right)$.

\subsection{Position polymerization}

We now proceed as above, but considering the $R_1\rightarrow\infty$ limit. 
The symplectic form associated to a spherical phase space is unchanged, and so 
is the corresponding Poisson bracket, while the spin components reduce to
\begin{align}
S_x&=S\cos\left(\frac{q}{R_2}\right),\\
S_y&=pR_2\cos\left(\frac{q}{R_2}\right),\\
S_z&=-S\sin\left(\frac{q}{R_2}\right).
\end{align}
The $\mathfrak{so(3)}$ algebra reduces to 
\begin{align} 
\left\{ \sin\left(\frac{q}{R_2}\right),\cos\left(\frac{q}{R_2}\right) \right\} &=0, \\
\left\{ R_2 \sin\left(\frac{q}{R_2}\right), p \right\} &=1, \label{BracketR2qp}\\
\left\{\cos\left(\frac{q}{R_2}\right),p \right\} &=-\frac{1}{R_2}\tan\left(\frac{q}{R_2}\right),
\end{align}
which is different from the cylindrical algebra because of the presence of the 
nonvanishing $\cos\left(q/R_2\right)$ factor in the symplectic form (\ref{S2form4}). However, 
if instead of the $q$ variable one considers $q_S$ given by Eq. (\ref{defqS}), 
the symplectic 2-form (\ref{S2form4}) can be expressed in a Darboux form
\begin{equation}
\omega = dp \wedge dq_S,
\end{equation}   
and we obtain $\left\{ q_S, p \right\} =1$, which agrees with (\ref{BracketR2qp}).

Rewriting the Hamiltonian $H_S$ (\ref{HamiltonianSphere}) under this limit gives
\begin{equation}
H_{C}=N \left(R_2 \sin\left(\frac{q}{R_2}\right) \right)\left[-\frac{3}{4}\kappa p^2 \cos^2\left(\frac{q}{R_2}\right)
+\frac{\Lambda}{\kappa}\right].
\label{HamiltonianR1Lim}
\end{equation}

Regarding the Friedmann equation, we can put Eq. (\ref{HubbleSphere}) under the form
\begin{align}
H^2 = \frac{\Lambda}{3}\left(\frac{\sin(q/R_2)}{q/R_2} \right)^2. 
\end{align}
An exact solution to this equation can be found again,
\begin{equation}
q(t) = 2R_2 \arctan \left(C e^{\pm \sqrt{3 \Lambda}t} \right),
\end{equation} 
where $C$ is a constant of integration. 

\section{Cosmological evolution}
\label{CosmologicalEvolution}

\begin{figure}[ht!]
\centering
\includegraphics[width=8cm, angle = 0]{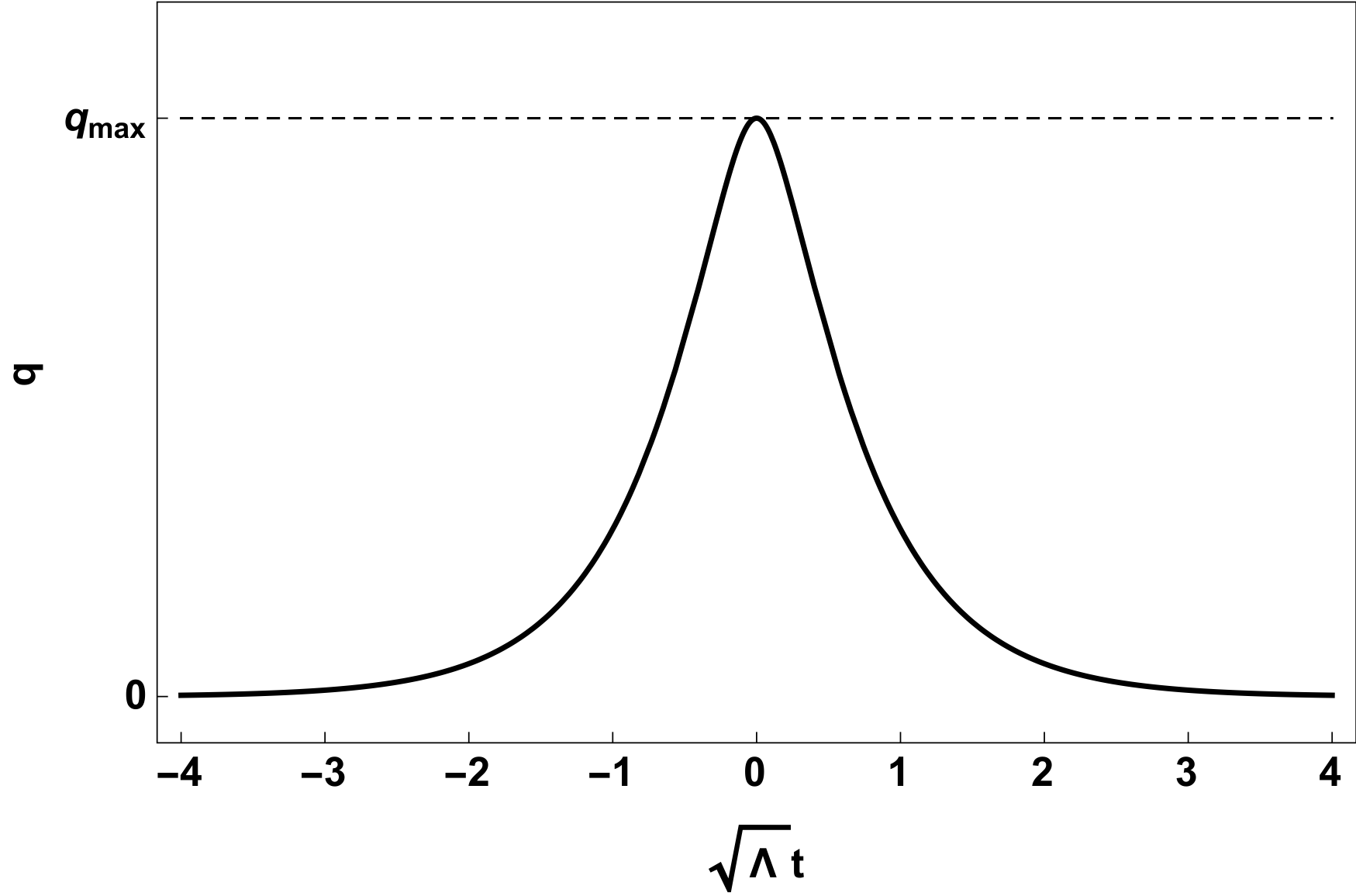}
\caption{Evolution of the canonical variable $q$.}
\label{Evolqvar}
\end{figure}

The aim of this section is to study the consequences of the modified Friedmann equation 
(\ref{FriedmannSphere}) obtained in the case of the spherical phase space. The equation can 
be rewritten in the form of the integral 
\begin{equation}
\int \frac{\cos(x)dx}{\sin(x)\sqrt{\cos^2(x)-\delta}}= \pm \sqrt{3 \Lambda}  t  + c,
\end{equation}
where $x:=q/R_2 \in [-\pi/2,\pi/2]$. The integral can be performed leading to 
\begin{align}
\cos(2x)&=2\left(1-\delta \right)\tanh^2\left(\mp \sqrt{3 \Lambda_{\text{eff}}} (t-t_0)\right) \nonumber \\
            &+2\delta-1,
\label{SolFriedS2ND}
\end{align}
where $t_0$ is a constant of integration that, without loss of generality, can be fixed 
as $t_0=0$. The solution is then symmetric with respect to the mirror symmetry $t\rightarrow -t$ 
and by examining the limits of Eq. (\ref{SolFriedS2ND}) we find that in the $t\rightarrow \pm \infty$ 
limits the value of $q$ tends to its minimal value,  
\begin{equation}
q_{\text{min}} := 0. 
\end{equation}
On the other hand, at $t=0$ the value of $q$ reaches $q_{\text{max}}$ (or symmetrically $-q_{\text{max}}$), where 
\begin{equation}
q_{\text{max}} := \frac{R_2}{2}\arccos\left(2\delta-1\right).
\label{Eqqmax}
\end{equation}
There are two symmetric branches of solution, related by the mirror symmetry 
$q \rightarrow - q$. The solution has a form of recollapsing universe with asymptotical de 
Sitter solutions at $t\rightarrow \pm \infty$ with the effective cosmological constant 
$\Lambda_{\text{eff}}$ [given by Eq. (\ref{deSitterEff})]. We present the solution in 
Fig. \ref{Evolqvar}. A qualitatively similar solution has been studied in the context
of quantum reduced loop gravity \cite{Cianfrani:2015oha}.

\begin{figure}[ht!]
\centering
\includegraphics[width=8cm, angle = 0]{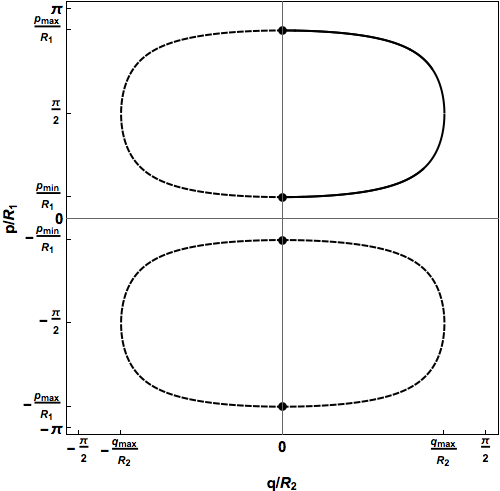}
\caption{Phase trajectories for the system under consideration.}
\label{fig:PhasePortrait}
\end{figure}

In Fig. \ref{fig:PhasePortrait} we show phase trajectories corresponding to the 
dynamics. Because the kinematical phase space is two-dimensional, by imposing 
the constraint the physical subspace is just one-dimensional subspace (curve).   
In this figure, the solid line represents the trajectory associated with the solution
shown in Fig. \ref{Evolqvar} and positive values of $p$.

The other three symmetric solutions, obtained by applying the reflections $q \rightarrow - q$
and $p \rightarrow - p$, are shown as the dashed lines in Fig. \ref{fig:PhasePortrait} (for the same value of $\delta$). 
The black dots represent beginning and ends to the trajectories, corresponding to times $t\rightarrow \pm \infty$. 
They are located at the equator of the phase space, where $\theta=\pi/2$.

While the $q$ reaches its maximal value, the canonically conjugated variable tends to either
$p = \frac{\pi}{2} R_1$ or (for a symmetric solution) to $p = -\frac{\pi}{2}R_1$. The minimal 
positive value of $p$, associated with the ends of the trajectory is 
\begin{equation}
p_{\text{min}} =R_1 \arcsin\left(\sqrt{\delta}\right), 
\end{equation}
and the maximal value is 
\begin{equation}
p_{\text{max}} = \pi R_1 - p_{\text{min}}.
\end{equation}

Based on Eq. (\ref{FriedmannSphere}), the leading  $q$-dependent correction 
to the Friedmann equation is
\begin{equation}
H^2 = \frac{\Lambda_{\text{eff}}}{3}-\frac{\Lambda}{9}\left(1+2\delta\right)\left(\frac{q}{R_2}\right)^2 
+\mathcal{O}(q^{4}).  
\end{equation}
Because $|q|:= V_0 a^3$, the leading correction scales as $a^6$. We can define an 
effective energy density,
\begin{equation}
\rho_{\text{eff}}:=\rho^{\text{eff}}_{\Lambda}+\rho_S 
= \frac{\Lambda_{\text{eff}}}{\kappa} -\frac{\Lambda}{3 \kappa}\left(1+2\delta\right)\left(\frac{q}{R_2}\right)^2. 
\end{equation} 
Using the continuity equation, 
\begin{equation}
\frac{d}{dt}\rho_{\text{eff}}+3H(\rho_{\text{eff}}+P_{\text{eff}})=0,
\end{equation}
we find that the Universe behaves effectively as filled by a fluid with effective pressure
\begin{equation}
P_{\text{eff}} := P^{\text{eff}}_{\Lambda}+P_S. 
\end{equation}
The effective equation of state implied by the compact nature of the phase space is
\begin{equation}
P_S = -3 \rho_S,
\end{equation}
where the equation of state for the effective cosmological constant remains in the 
classical form $P^{\text{eff}}_{\Lambda} = - \rho^{\text{eff}}_{\Lambda}$. The 
barotropic index for the correction is $w_S=-3<-1$, which can be interpreted as a special kind of the phantom matter \cite{Caldwell:1999ew}.  
 
\subsection{Evolution of the vector $\vec{S}$}

Let us now discuss evolution of the three spin components $S_i$. Because of 
the scalar constraint (\ref{Lambda}), which can be written in the form 
\begin{equation}
\frac{S_y}{S} = \pm \sqrt{\delta} = \text{const},   
\end{equation}
the equations of motion (\ref{dotSx}-\ref{dotSz}) (on the surface of the constraint) 
can be written as
\begin{align}
\dot{S}_x&=\frac{3}{2}\frac{N\kappa}{R_1 R_2^2}S_yS_z^2,\\
\dot{S}_y&=0, \\
\dot{S}_z&=-\frac{3}{2}\frac{N\kappa}{R_1 R_2^2} S_x S_y S_z,
\end{align}
with $S_y=$ const. Because of the equation of the sphere $S_x^2+S_y^2+S_z^2=S^2$, 
only one independent differential equation remains,
\begin{equation}
\dot{S_x}=-\alpha S_x^2+\alpha \left(S^2-S_y^2\right), \label{diffEqSx}
\end{equation}
where $\alpha:=\frac{3}{2}N \frac{\kappa}{R_2}\frac{S_y}{S}$. The analytical solution to Eq. (\ref{diffEqSx})
can be found leading to 
\begin{align}
S_x(t)&= \sqrt{S^2-S_y^2}\tanh\left(\alpha\sqrt{S^2-S_y^2}\ (t-t_0)\right),\\
S_y(t)&= \pm S \sqrt{\delta}, \\ 
S_z(t)&=\pm \frac{\sqrt{S^2-S_y^2}}{\cosh\left(\alpha\sqrt{S^2-S_y^2}\ (t-t_0)\right)}.
\end{align}

\begin{figure}[h]
\includegraphics[width=8cm, angle = 0]{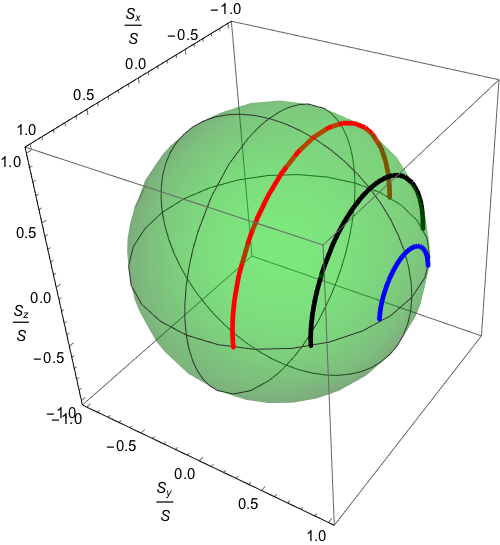}
\caption{Evolution of the vector $\vec{S}=(S_x,S_y,S_x)$ for (from left to right) 
$\delta=0.1$ (red thick curve), $\delta=0.5$ (black thick curve),  $\delta=0.9$ 
(blue thick curve). There are symmetric solutions in the three other quadrants
of the sphere. The $S_y$ component remains constant during the evolution and 
is given by $S_y=S\sqrt{\delta}$. The $t\rightarrow \pm \infty$ limits are located at 
the ends of the curves (on the equator). The maximal value of the $S_z$ component
corresponds to the maximal value of the $q$ variable that is reached on a trajectory.}
\label{fig:SEvol}
\end{figure}

There are four symmetric solutions for the four quadrants of a sphere. In the solution with 
two pluses, as time flows, the $S_x$ component grows from $-\sqrt{S^2-S_y^2}$
to $\sqrt{S^2-S_y^2}$, while the $S_y$ remains constant. The $S_z$ component 
grows from $0$ to its maximal value $\sqrt{S^2-S_y^2}$ and then is decreasing 
again to $0$. 

\section{Quantum theory}
\label{QuantumConsiderations}

In the quantum theory the Hamiltonian (\ref{HamiltonianSphere}) is promoted to an operator 
$\hat{H}_{S}=N \hat{C}$. The appropriately symmetrized\footnote{For simplicity, we do not 
decompose $\hat{S}^2$ and do not include factorial powers of the operators $\hat{S}_i$ in the 
symmetrization.} and normalized constraint can be written as
\begin{widetext}
\begin{equation}
\hat{c} := \frac{4S^2}{3 \kappa R_1} \hat{C}
=\frac{1}{3}\left( \hat{S}_z\hat{S}_y\hat{S}_y+\hat{S}_y\hat{S}_z\hat{S}_y
+\hat{S}_y\hat{S}_y\hat{S}_z \right)-\delta \hat{S}^2\hat{S}_z \approx 0. 
\label{QuantHamiltonianSphere}
\end{equation}
\end{widetext}

The task is now to find the physical states $| \Psi_{\text{phys}} \rangle$ belonging to the kernel 
of the operator, i.e., $ \hat{C}| \Psi_{\text{phys}} \rangle=0$.

A convenient method to deal with constrained systems is the group averaging \cite{Ashtekar:1995zh,
Giulini:1999kc}. We introduce the projection operator (see, e.g., \cite{Rovelli:2009tp}),
\begin{equation}
\hat{P} := \lim_{T\rightarrow \infty } \frac{1}{2T} \int_{-T}^{T}d\tau  e^{i\tau \hat{C}},
\end{equation} 
which projects to the zero eigenvalue of the operator $\hat{C}$. The operator $\hat{P}$ 
projects quantum states onto the physical subspace and satisfies the condition $\hat{P}^2=\hat{P}$.

We are mostly interested in the large spin limit ($R_{1,2}\rightarrow \infty$) in which the 
semiclassical cosmological behavior can be reconstructed. In that case, the area of the phase 
space is large and the $q$ and $p$ variables are well defined. 

It is important to stress that there are two ways of looking at the minisuperspace model. The first is 
that the model provides description of the Universe at the largest possible (cosmological) scale, 
describing averaged degrees of freedom. However, there is also a second interpretation, which is 
especially interesting in the Belinsky-Khalatnikov-Lifschitz \cite{Belinsky:1970ew,Belinsky:1982pk} 
limit characterized by decoupling off the space points. In this case, evolution at each space point 
is described by a homogeneous (in general anisotropic) minisuperspace model. Furthermore, 
by introducing interactions between the minisuperspace models at points, one can try to study spatial 
properties of the field configuration \cite{Baytas:2016cbs}. From this perspective, the small spin 
case can have relevance for the very early Universe, before the semiclassical regime has been 
entered. Therefore, let us start our consideration with the simplest case of spin $1/2$.  

\subsection{Spin 1/2}

In this case, the spin components are expressed as
\begin{equation}
S_i = \frac{\hslash}{2} \sigma_i 
\end{equation}
where $\sigma_i$ are Pauli matrices and $i=x,y,z$. Therefore, $\hat{S}^2=\frac{3\hbar^2}{4}\mathbb{1}$ 
and the spherical phase space has an area operator $\hat{A}=2\pi\sqrt{3}\hbar \mathbb{1}$. The 
scalar constraint (\ref{QuantHamiltonianSphere}) reduces to
\begin{align}
\hat{c} &= \frac{1}{3}\left(1-9\delta \right)\hat{S}_z^3 = \beta \sigma_z 
\label{c12}
\end{align}
defining $\beta=\frac{1}{24}\left(1-9 \delta \right)\hbar^3$. We directly conclude 
that there are no physical states in this case: dim$(\mathcal{H}_{\text{phys}})=0$.  
This is because $ \hat{C}| \Psi_{\text{phys}} \rangle \sim \sigma_z | \Psi_{\text{phys}} \rangle$, 
where $\sigma_z=\text{diag}(1,-1)$ does not have nontrivial solutions. In agreement 
with this, the projector operator $\hat{P}$ is null,
\begin{equation}
\hat{P} := \lim_{T\rightarrow \infty } \frac{\sin\left(\beta T\right)}{\beta T} \mathbb{1} = 0.
\end{equation} 
\\

\subsection{General spin $s$}

Let us now proceed to the general quantum number $s$. In order to find matrix elements 
of the constraint $\hat{c}$ in the basis $|s,m\rangle$, where $m=-s,\dots,s$, it is useful 
to introduce spin ladder operators,
\begin{align}
\hat{S}_{\pm} = \hat{S}_x\pm i \hat{S}_y. 
\end{align}
The action of the relevant spin operators on the states $|s,m\rangle$ is given as 
follows:
\begin{align}
\hat{S}^2 |s,m \rangle &= s(s+1) \hslash^2|s,m \rangle, \\
\hat{S}_z|s,m \rangle &= m \hslash |s,m \rangle, \\
\hat{S}_{\pm}|s,m \rangle &= \sqrt{s(s+1)-m(m\pm1)} \hslash |s,m\pm1\rangle.
\end{align}
And from Eq. (\ref{S_z}), we see that the eigenvalues $m$ correspond to eigenvalues of the quantum 3-volume. Applying the above formulas to Eq. (\ref{QuantHamiltonianSphere}), we find the 
matrix elements
\begin{equation}
\langle s, m'|\hat{c}|s,m\rangle = c_1^m \delta_{m',m+2} + c_2^m \delta_{m',m} + c_3^m \delta_{m',m-2}
\end{equation}
where $\delta$ denotes the Kronecker delta, and the $c_i^m$ coefficients are defined as
\begin{widetext}
\begin{equation}
\left\{
  \begin{array}{l c l}
c_1^m =-\frac{\hbar^3}{4}(m+1)\sqrt{\left(s(s+1)-m(m+1)\right)\left(s(s+1)-(m+1)(m+2)\right)}\\
c_2^m =\hbar^3m\left(\frac{1}{2}\left(s(s+1)-m^2\right)-\frac{1}{6}-\delta s(s+1)\right)\\
c_3^m =-\frac{\hbar^3}{4}(m-1)\sqrt{\left(s(s+1)-m(m-1)\right)\left(s(s+1)-(m-1)(m-2)\right)}
  \end{array}
   \right.
\end{equation}
\end{widetext}
where we notice the following properties:
\begin{equation}
\left\{
  \begin{array}{l c l}
c_1^m =-c_3^{-m}\\
c_2^m =-c_2^{-m}\\
c_3^{m+2}=c_1^m
  \end{array}
   \right.
\label{cProperties}
\end{equation}
   
And $\hat{c}$ reduces then to an antipersymmetric Hankel matrix with only three nonzero diagonals.

For the particular case $s=1/2$, $c_1^m$ and $c_3^m$ are not defined, and 
$\langle s, m'|\hat{c}|s,m\rangle=\frac{m}{12} \left(1-9 \delta \right)\hbar^3\delta_{m',m}$, 
which is in agreement with Eq. (\ref{c12}).

An important property of the matrix 
\begin{equation}
c_{m',m}:=\langle s, m'|\hat{c}|s,m\rangle 
\end{equation}
is that its determinant is always equal to $0$ only for bosonic representations ($s \in \mathbb{N}_{+}$). 
For the fermionic representations with $s$ half integers ($2s \in \mathbb{N}_{+}$) the determinant is   
some function of $\delta$ and can be equal to $0$ only for some special values of $\delta$. 
As a consequence, for arbitrary $\delta$ there are nontrivial vectors belonging to the kernel of the 
$c_{m',m}$ matrix only in the bosonic case. Furthermore, the matrix is symmetric and therefore 
has real eigenvalues. 

\subsection{Solving the quantum constraint}

The general solutions of the constraint  equation
\begin{equation}
\hat{c}|\Psi \rangle = 0
\end{equation}
can be expressed in terms of the basis states as follows:
\begin{equation}
|\Psi \rangle  = \sum_{m=-s}^{s} a_m |s,m\rangle, 
\end{equation}
together with the normalization condition $\sum_{m=-s}^{s} |a_m|^2=1$. With the 
use of the matrix elements, we obtain the following recursive equation
\begin{equation}
a_{m-2}c_1^{m-2}+a_{m}c_2^m+a_{m+2}c_3^{m+2}=0,
\label{RecurvEqn} 
\end{equation}
for $m=-s+2,-s+3,\dots,s-3,s-2$, together with 
\begin{align}
a_{s-2}c_1^{s-2}+a_sc_2^s &= 0, \\
a_{s-3}c_1^{s-3}+a_{s-1}c_2^{s-1} &= 0, \\
a_{-s+3}c_3^{-s+3}+a_{-s+1}c_2^{-s+1} &= 0, \\
a_{-s+2}c_3^{-s+2}+a_{-s}c_2^{-s} &= 0. 
\end{align}

Combining these conditions together and using Eq. (\ref{RecurvEqn}), 
we get the following conditions:
\begin{align}
a_{-s} a_{s-2}=a_s a_{-s+2}, \\
a_{-s} a_{s-4}=a_s a_{-s+4}, \\
a_{-s+1} a_{s-3}=a_{s-1} a_{-s+3}, \\
a_{-s+1} a_{s-5}=a_{s-1} a_{-s+5}. 
\end{align}

We can now prove by recurrence that, for any $m\in[-s+2,s-2]$,
\begin{align}
a_{-s} a_{m}=a_s a_{-m}, \\
a_{-s+1} a_{m}=a_{s-1} a_{-m}, 
\end{align}
which allows us, using another recurrence, to show that
\begin{itemize}
	\item for the bosonic case $s\in \mathbb{N}_{+}$, $\forall m\in[-s,s]$,
\begin{equation}
a_m=a_{-m}
\label{BosonState}
\end{equation}
	\item for the fermionic case $2s\in \mathbb{N}_{+}$, $\forall m\in[-s,s]$,
\begin{equation}
a_m=\pm a_{-m}
\label{FermionState}
\end{equation}
\end{itemize}

Let $\hat{c}$ be an arbitrary $N\times N$ matrix with $N=2k+1$ for the 
bosonic case or $N=2k$ for the fermionic one, $k\in \mathbb{N}_+$. The matrix 
determinant can be represented as
\begin{equation} 
\det{\hat{c}} =
\begin{vmatrix}
c_2^{-s} & 0 & c_3^{-s+2} & 0  \\
0 & c_2^{-s+1} & 0 & \ddots \\
c_1^{-s} & 0 & c_2^{-s+2} & 0  \\
0 & \ddots & 0 & \ddots 
\end{vmatrix}
\end{equation}

By property of the $N$ linear alternating form of the determinant, we can exchange columns 
by pairs and do the same with rows to rewrite
\begin{equation} 
\det{\hat{c}} =(-1)^{2k}
\begin{vmatrix}
c_2^{s} & 0 & c_1^{s-2} & 0  \\
0 & c_2^{s-1} & 0 & \ddots \\
c_3^{s} & 0 & c_2^{s-2} & 0  \\
0 & \ddots & 0 & \ddots 
\end{vmatrix}
\end{equation}
Using the properties of Eq. (\ref{cProperties}), we finally get
\begin{equation} 
\det{\hat{c}} =\det(-\mathbb{1}_N)\det{\hat{c}} =(-1)^N\det{\hat{c}}. 
\end{equation}
Therefore, in the bosonic case (odd $N$), the determinant is always $0$, ensuring that 
at least one eigenvalue is equal to $0$.

From this we conclude that for bosons, in the eigenbasis of $\hat{c}$ the projection operator 
can be written as
\begin{equation}
\text{diag}(\hat{P})=(1,...,1,0,...,0)
\end{equation}
where we have the $1$ components originating from each (degenerated) eigenvalue equal to 
$0$ in $\hat{c}$, whereas any eigenvalue $\lambda_i\neq 0$ implies a component 
$\lim_{T\rightarrow \infty} \frac{\sin(\lambda_i T)}{\lambda_i T} =0 $ in the diagonalized projection 
operator. Therefore, as long as the projection operator is non-$0$, there exists a nontrivial 
solution to equation $\hat{c}|\Psi \rangle = 0$. 

In the fermionic case, nontrivial solutions may exist only for certain values of the parameter 
$\delta$. In particular, for $s=1/2$ there are no nontrivial solutions (except in the case 
$\delta=\frac{1}{9}$ for which the constraint is identically equal to $0$), for $s=3/2$ nontrivial a
solution exists for $\delta=\frac{13}{45}$, and for $s=5/2$ there are two values for which the 
determinant of $\hat{c}$ is vanishing: $\delta=\frac{7}{15}$ and  $\delta=\frac{1}{105}$. 
Every further case has to be investigated individually. 
 
\subsection{Example: $s=2$}

As an example of the solution of the constraint let us examine the case of $s=2$.
In this case the constraint $\hat{c}$ takes the following matrix form: 
\begin{equation} 
{\bf c} = \hslash^3
\left(
\begin{array}{ccccc}
 \frac{5}{3}-12 \delta  & 0 & -\sqrt{\frac{3}{2}} & 0 & 0 \\
 0 & \frac{7}{3}-6 \delta  & 0 & 0 & 0 \\
 -\sqrt{\frac{3}{2}} & 0 & 0 & 0 & \sqrt{\frac{3}{2}} \\
 0 & 0 & 0 & 6 \delta -\frac{7}{3} & 0 \\
 0 & 0 & \sqrt{\frac{3}{2}} & 0 & 12 \delta -\frac{5}{3} \\
\end{array}
\right).  
\label{cs2matrix}
\end{equation}
For $\delta\neq \frac{7}{18}$, there is a single state that satisfies the constraint and is given  by 
\begin{equation}
| \Psi_{\text{phys}} \rangle=  a_{-2} |2,-2\rangle+a_0 |2,0\rangle+a_2 |2,2\rangle,
\label{PhysicalStateS2}
\end{equation}
where appropriately normalized coefficients are 
\begin{align}
a_2 = a_{-2} &= \frac{3}{2} \sqrt{\frac{3}{2}}  \frac{1}{\sqrt{324 \delta ^2-90 \delta +13}}, \\
a_0 &= \frac{1}{2} \frac{5-36\delta}{\sqrt{324 \delta ^2-90 \delta +13}}.
\end{align} 

On the other hand, for $\delta = \frac{7}{18}$, the constraint $\hat{c}$ has three linearly independent solutions:
\begin{align}
| \Psi_{\text{phys},1} \rangle &=  \frac{1}{2\sqrt{2}} |2,-2\rangle-\frac{\sqrt{3}}{2}|2,0\rangle+ \frac{1}{2\sqrt{2}} |2,2\rangle, \label{7/18-1} \\
| \Psi_{\text{phys},2} \rangle &= |2,1\rangle,  \label{7/18-2} \\
| \Psi_{\text{phys},3} \rangle &= |2,-1\rangle.  \label{7/18-3}
\end{align}

The matrix  (\ref{cs2matrix}) can be diagonalized so that ${\bf c} = {\bf U^{-1}} {\bf \Lambda} {\bf U}$, 
where the diagonal matrix ${\bf \Lambda} = \text{diag}(\lambda_1,\lambda_2,\lambda_3,\lambda_4,\lambda_5)$,
where the eigenvalues
\begin{align}
\lambda_1 &= 0, \\
\lambda_2 &={\hslash^3} \frac{1}{3} (7-18 \delta ), \\
\lambda_3 &={\hslash^3} \frac{1}{3} (18 \delta -7), \\
\lambda_4 &=-{\hslash^3}\frac{2}{3} \sqrt{324 \delta ^2-90 \delta +13}, \\
\lambda_5 &={\hslash^3} \frac{2}{3} \sqrt{324 \delta ^2-90 \delta +13}. \\
\end{align}

With the use of this, the exponentiation of the constraint can be written as
\begin{equation}
 e^{i\tau {\bf c}} = {\bf U^{-1}}    e^{i\tau{\bf \Lambda}} {\bf U}
\end{equation}   

As a consequence, the projection operator takes the form 
\begin{align}
\hat{P} &= \lim_{T\rightarrow \infty} \frac{1}{2T} \int_{-T}^{T} d\tau  e^{i\tau {\bf c}}\\
 &= {\bf U^{-1}}
\left(
\begin{array}{ccccc}
 1 & 0 & 0 & 0 & 0 \\
 0 & 0 & 0 & 0 & 0 \\
 0 & 0 & 0 & 0 & 0 \\
 0 & 0 & 0 & 0 & 0 \\
 0 & 0 & 0 & 0 & 0
\end{array}
\right) {\bf U}
\end{align}
in the $\delta\neq \frac{7}{18}$ case. The factor $1$ in the middle matrix comes directly 
from the $\lambda_1=0$ eigenvalue, whereas any eigenvalue $\lambda_i\neq 0$ implies 
a component $\lim_{T\rightarrow \infty } \frac{\sin(\lambda_i T)}{\lambda_i T} =0 $ in the 
projection operator. 

Using the expression of ${\bf U}$, the projection operator finally reads
\begin{equation}
\hat{P} =\left(
\begin{array}{ccccc}
 P_1 & 0 & P_2 & 0 & P_1 \\
 0 & 0 & 0 & 0 & 0 \\
 P_2 & 0 & P_3 & 0 & P_2 \\
 0 & 0 & 0 & 0 & 0 \\
 P_1 & 0 & P_2 & 0 & P_1
\end{array}
\right)
\label{P}
\end{equation}
in the $\delta\neq \frac{7}{18}$ case, where we defined
\begin{align}
P_1 &:=\frac{27}{8(324\delta^2-90\delta+13)}=a_2^2, \\
P_2 &:=\frac{3}{4}\sqrt{\frac{3}{2}}\frac{5-36\delta}{324\delta^2-90\delta+13}=a_0a_2, \\
P_3 &:= \frac{(5-36\delta)^2}{4(324\delta^2-90\delta+13)}= a_0^2. 
\end{align}

The form of the projection operator we have obtained agrees with the expression 
given as a dyadic product of the physical states $| \Psi_{\text{phys}} \rangle$, i.e., 
\begin{equation}
\hat{P} = | \Psi_{\text{phys}} \rangle \langle \Psi_{\text{phys}} |
\end{equation} 
taking the physical state given by Eq. (\ref{PhysicalStateS2}). The property $\hat{P}^2=
| \Psi_{\text{phys}} \rangle \langle \Psi_{\text{phys}} | \Psi_{\text{phys}} \rangle \langle \Psi_{\text{phys}} |
=| \Psi_{\text{phys}} \rangle \langle \Psi_{\text{phys}} |=\hat{P}$ is satisfied in a consequence 
of the proper normalization of the state $| \Psi_{\text{phys}} \rangle$.

If we consider the degenerated case with $\delta=\frac{7}{18}$, for which 
$\lambda_1=\lambda_2=\lambda_3=0$, the projection operator reads
\begin{align}
\hat{P} &= {\bf U^{-1}}
\left(
\begin{array}{ccccc}
 1 & 0 & 0 & 0 & 0 \\
 0 & 1 & 0 & 0 & 0 \\
 0 & 0 & 1 & 0 & 0 \\
 0 & 0 & 0 & 0 & 0 \\
 0 & 0 & 0 & 0 & 0
\end{array}
\right) {\bf U}.  \nonumber \\
&= \left(
\begin{array}{ccccc}
 \frac{1}{8} & 0 & -\frac{1}{4}\sqrt{\frac{3}{2}} & 0 & \frac{1}{8} \\
 0 & 1 & 0 & 0 & 0 \\
 -\frac{1}{4}\sqrt{\frac{3}{2}} & 0 & \frac{3}{4} & 0 & -\frac{1}{4}\sqrt{\frac{3}{2}} \\
 0 & 0 & 0 & 1 & 0 \\
 \frac{1}{8} & 0 & -\frac{1}{4}\sqrt{\frac{3}{2}} & 0 & \frac{1}{8}
\end{array}
\right) \nonumber  \\
& = \sum_{i=1}^3 | \Psi_{\text{phys},i} \rangle \langle  \Psi_{\text{phys},i}|,
\label{P7/18}
\end{align}
where the states $| \Psi_{\text{phys},i} \rangle$ for $i=1,2,3$ are given by Eqs. 
(\ref{7/18-1})-(\ref{7/18-3}). 
 
Let's now take arbitrary $\delta$ and consider sequences from an initial state $ |m \rangle := |s,m \rangle $ to a 
final state $|n \rangle$, together with the transition amplitudes $\langle n | \hat{P} |m \rangle$.
From the expressions of (\ref{P}) we directly observe that states $|-1 \rangle$ and 
$|1 \rangle$ do not belong to the sequence if $\delta \neq 7/18$. For the basis states, the 
transition amplitudes $\langle n | \hat{P} |m \rangle$ are
\begin{itemize}
	\item $P_1$ each time we transit from $|m \rangle$ to $|n \rangle$ 
	where $m,n=-2,2$ (i.e. the state is possibly unchanged).
	\item $P_2$ each time we transit from $|-2 \rangle$ to $|0 \rangle$ 
	and vice versa, or from $|2 \rangle$ to $|0 \rangle$ and vice versa.
	\item $P_3$ each time the state $|0 \rangle$ is unchanged.
	\item $1$ each time the states $|-1 \rangle$ and $|1 \rangle$ remain unchanged for the $\delta=7/18$ case.
\end{itemize}

In Sec. \ref{CosmologicalEvolution} we have shown that the cosmological 
evolution from $t\rightarrow -\infty$ to $t\rightarrow +\infty$ is associated with the rotation
of the spin vector $\vec{S}$ from the position
\begin{equation}
\vec{S}_{\text{in}} = S (-\sqrt{1-\delta}, \sqrt{\delta},0)
\label{Sin}
\end{equation}
to 
\begin{equation}
\vec{S}_{\text{out}} = S (+\sqrt{1-\delta}, \sqrt{\delta},0).
\label{Sout}
\end{equation}
See the bolded trajectories in Fig. \ref{fig:SEvol}. There are also symmetric solutions in 
the three other quarts of the sphere, but let us now focus on this representative one. Namely, 
we want to find what is the quantum transition amplitude associated with this 
evolution as a function of the parameter $\delta$,
\begin{equation}
W(\delta) :=  \langle \text{out}|\hat{P}|\text{in}\rangle, 
\label{TransitionAmplitudeW}
\end{equation} 
for the $s=2$ case, considered in this section. At the level of quantum mechanics, the 
boundary conditions (\ref{Sin}) and (\ref{Sout}) translate into the properties of the mean 
values of the components of the spin operator $\hat{S}$ in the initial and final states, i.e. 
\begin{align}
\langle \text{in}|\hat{S}_x|\text{in}\rangle &=-S\sqrt{1-\delta}, \\
\langle \text{in}|\hat{S}_y|\text{in}\rangle &=S\sqrt{\delta},  \\
\langle \text{in}|\hat{S}_z|\text{in}\rangle &=0,
\end{align}
and 
\begin{align}
\langle \text{out}|\hat{S}_x|\text{out}\rangle &=S\sqrt{1-\delta}, \\
\langle \text{out}|\hat{S}_y|\text{out}\rangle &=S\sqrt{\delta},  \\
\langle \text{out}|\hat{S}_z|\text{out}\rangle &=0,
\end{align}
where for the case considered, $S=2\hslash$. Furthermore, it is assumed 
that dispersion relations are minimized, which is satisfied by $SU(2)$ 
coherent states. These can be obtained by rotations of the $|2,2\rangle$ state 
into directions of the vectors (\ref{Sin}) and (\ref{Sout}). An appropriate rotation 
operator takes the form 
\begin{equation}
\hat{R}(\phi,\theta,0) =e^{-\frac{i}{\hslash}\phi \hat{S}_z}e^{-\frac{i}{\hslash}\theta \hat{S}_y},        
\end{equation}
where $\phi $ and $\theta$ are the spherical angles introduced in Sec. \ref{SphericalPS}. 
What is to be performed is the rotation of the state $|2,2\rangle$ (for which the spin vector  
is precessing around the $z$ axis) first by angle $\theta = \pi/2$ around $y$ axis (using the 
operator $e^{-\frac{i}{\hslash}\theta \hat{S}_y}$). This aligns the vector to the $x-y$ plane. 
Then we rotate the vector around the $z$ axis  (using the operator 
$e^{-\frac{i}{\hslash}\phi \hat{S}_z}$) to the initial and final orientations of the spin vector, 
with $\phi_{\text{in}}$ and $\phi_{\text{out}}$. For both initial and final state, we have  
$\theta_{\text{in}}=\theta_{\text{out}}=\pi/2$ and 
\begin{align}
\phi_{\text{out}} &= \arctan \left( \sqrt{\frac{\delta}{1-\delta}}\right),\\
\phi_{\text{in}} &=\pi -\phi_{\text{out}}.    
\end{align}
Using this, the initial and final states can be written as
\begin{align}
|\text{in/out}\rangle &= \hat{R}(\phi_{\text{in/out}},\pi/2,0) |2,2\rangle \nonumber \\
                               &=\sum_{m=-2}^2 b_m(\phi_{\text{in/out}},\pi/2,0)|2,m\rangle,    
\end{align}
where the coefficients 
\begin{align}
b_m(\phi_{\text{in/out}},\pi/2,0) &= \langle 2,m |\hat{R}(\phi_{\text{in/out}},\pi/2,0)|2,2\rangle \nonumber \\
                                                 &=e^{-im\phi_{\text{in/out}}} d^2_{m2}(\pi/2), 
\end{align}
where $d^2_{m2}(\pi/2)$ are components of the Wigner $d$-matrices, 
\begin{align}
d^2_{22}(\pi/2) &=\frac{1}{4}, \\
d^2_{12}(\pi/2) &= \frac{1}{2},\\
d^2_{02}(\pi/2) &= \sqrt{\frac{3}{8}},\\
d^2_{-12}(\pi/2) &= \frac{1}{2},\\
d^2_{-22}(\pi/2) &= \frac{1}{4}.
\end{align}

With the use of these results, we obtain the following initial and final states: 
\begin{equation}
\begin{split}
|\text{in}\rangle =
 \left(\frac{1}{4} -\frac{\delta}{2} -\frac{i}{2}\sqrt{\delta(1-\delta)}\right)  |2,-2\rangle \\
+ \left(-\frac{1}{2}\sqrt{1-\delta}+\frac{i}{2}\sqrt{\delta} \right)  |2,-1\rangle \\
+\sqrt{\frac{3}{8}} |2,0\rangle \\
+ \left( -\frac{1}{2}\sqrt{1-\delta}-\frac{i}{2}\sqrt{\delta}\right) |2,1\rangle \\
+ \left( \frac{1}{4} -\frac{\delta}{2} +\frac{i}{2}\sqrt{\delta(1-\delta)} \right)|2,2\rangle 
\end{split}
\end{equation}
and 
\begin{equation}
\begin{split}
|\text{out}\rangle = 
 \left(\frac{1}{4} -\frac{\delta}{2} +\frac{i}{2}\sqrt{\delta(1-\delta)}\right)  |2,-2\rangle \\
+ \left(\frac{1}{2}\sqrt{1-\delta}+\frac{i}{2}\sqrt{\delta} \right)  |2,-1\rangle \\
+ \sqrt{\frac{3}{8}} |2,0\rangle \\
+ \left( \frac{1}{2}\sqrt{1-\delta}-\frac{i}{2}\sqrt{\delta}\right) |2,1\rangle \\
+\left(\frac{1}{4} -\frac{\delta}{2} -\frac{i}{2}\sqrt{\delta(1-\delta)} \right)|2,2\rangle
\end{split}
\end{equation}

\begin{figure}[h!]
\includegraphics[width=8cm, angle = 0]{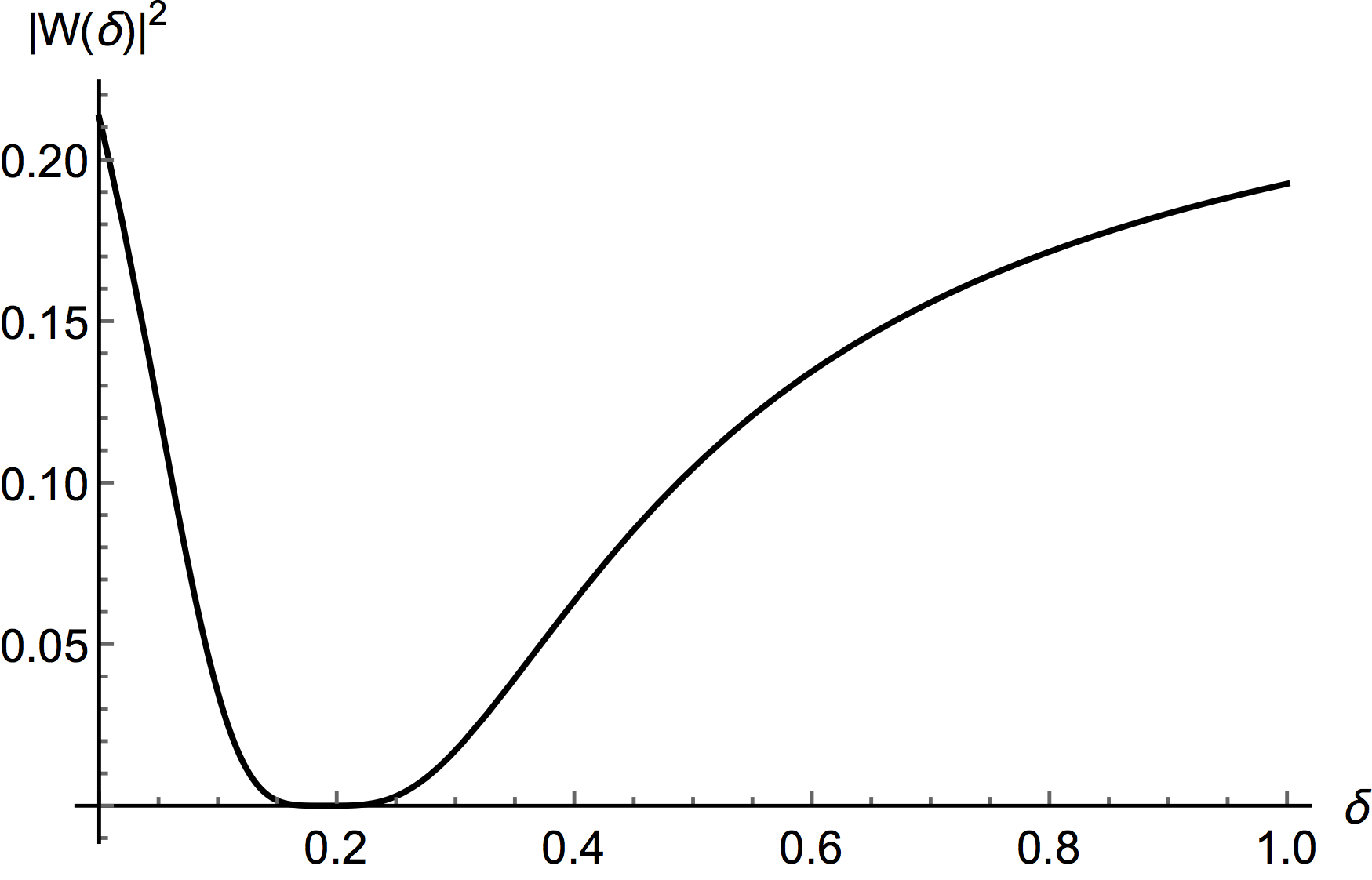}
\caption{Modulus square of the transition amplitude as a function of the parameter 
$\delta=\Lambda/\Lambda_*\in [0,1]$ and $\delta \neq \frac{7}{18}\approx 0.389$. 
The $|W(\delta)|^2$ decreases from $|W(0)|^2=(\frac{6}{13})^2 \approx 0.213$ to 
$|W(4/21)|^2=0$: then the probability of transition increases until reaching 
$|W(1)|^2=\left(\frac{867}{1976}\right)^2 \approx 0.193$. For the special case $\delta=7/18$
we have $|W\left(7/18\right)|^2\approx 0.017$.}
\label{fig:W}
\end{figure}

Then, it is straightforward to calculate the transition amplitude (\ref{TransitionAmplitudeW})
using the projection operator given by Eq. (\ref{P}) for $\delta \neq \frac{7}{18}$ and 
by Eq. (\ref{P7/18}) for $\delta = \frac{7}{18}$. For $\delta \neq \frac{7}{18}$, the final form of 
the transition amplitude is 
\begin{equation}
W(\delta) =  \frac{3 (4-21\delta)^2}{8 \left(324\delta^2-90\delta+13\right)},
\label{TransAmpFunc}
\end{equation}
which is a purely real function. In the special case $\delta=7/18$ the transition amplitude is 
$W\left(7/18\right)=\frac{337}{2592}\approx 0.130$. We plot the modulus square of the 
amplitude in Fig. \ref{fig:W}.
The probability of transition is maximized for $\Lambda=0$, where $|W(0)|^2=(\frac{6}{13})^2 
\approx 0.213$. Then, the probability is decreasing to $0$ at $\delta=\frac{4}{21}$ and is 
increasing to the value $|W(1)|^2=\left(\frac{867}{1976}\right)^2 \approx 0.193$ at the 
maximal value of $\delta$. It is interesting that in the limit of vanishing cosmological constant 
the probability takes the maximal value. 

Let us explore this feature in some more detail. For this purpose, we
recall that the probability amplitude (\ref{TransAmpFunc}) describes quantum 
transition between quantum states peaked at two end points of a classical 
trajectory (at $t\rightarrow \pm \infty$). In the considered case of $s=2$, the associated 
quantum dynamics is approximated by a five-level quantum system. This, however, does 
not mean that we are restricted to the Planckian regime only. The maximal volume of 
the Universe is determined by both $R_2$ and $\delta$ [see Eq. (\ref{Eqqmax})]. The 
$R_2$ can take arbitrary large values, provided that the relation $R_1R_2=S$ is satisfied. 

For $\delta \ll 1$ the transition amplitude (\ref{TransAmpFunc}) corresponds to the evolution 
with $q_{\text{max}} \approx  \frac{\pi}{2}R_2$, which may describe large scale universe. 
The fact that a low dimensional quantum system may have cosmological relevance not only 
at the Planck scale has already been used in spin-foam cosmology, where a dipole spin 
network provides an approximate large scale description of a universe \cite{Bianchi:2010zs,
Vidotto:2010kw}. However, in order to recover semiclassical behavior, large spin limit 
(of spin labels) has to be taken. Accordingly, in our case, correspondence with classical 
results is expected in the $s\rightarrow \infty$ limit (equivalent to the affine limit of a spherical 
phase space). An open question is whether the main features of the transition amplitude 
(\ref{TransAmpFunc}) are preserved in this limit. This, however, requires further analysis 
and cannot be simply answered based only on the studies made here. Worth emphasizing 
is also that determination of $W(\delta)$ for $s\gg 1$ may also allow one to reconstruct semiclassical 
action corresponding to the quantum system under consideration. It would be very interesting 
to check if the effective Friedmann equation (\ref{FriedmannSphere}) is recovered.

\section{Summary}
\label{SecSummary}

The research program of NFST aims at generalizing 
the known types of field theory to the case of compact phase space. This goal has 
already been achieved in the case of scalar field theory. 

The aim of this program is  the compactification of the phase space of gravity. Thanks 
to the compactness of the phase space, the field variables become constrained, 
which may eliminate singularities. The compactness of the phase space is a 
consequence of the finite dimensionality of the Hilbert space. Therefore, the compact 
phase space of gravitational field theory is expected to be associated with a quantum 
theory of gravity characterized by finite number of basis states. In this article, we 
have examined the possibility of compactifying a minisuperspace gravitational model 
with a single degree of freedom - a scale factor. 

We have focused our attention on the vacuum case with positive cosmological constant. 
The phase space has been generalized  from the affine case $\mathbb{R}^2$ to a
spherical phase space $\mathbb{S}^2$. This choice has been suggested by the fact 
that the spherical phase space is a phase space of angular momentum (spin). This 
enabled us to describe kinematics and dynamics of the model in terms of the angular 
momentum (spin) vector $\vec{S}$, the components of which satisfy the $\mathfrak{so(3)}$ 
($\mathfrak{su(2)}$) algebra. The affine case is recovered in the large spin limit 
$R_{1,2}\rightarrow \infty$. At the quantum level, the large spin limit $s\rightarrow\infty$ 
is known to be associated with semiclassical limit, which is consistent with our discussion. 

We have investigated semiclassical dynamics of the system and we have derived the 
modified Friedmann equation. The ambiguity associated with introduction of the compact 
phase space extension of the Hamiltonian has been reduced by requirement that the 
known case of loop quantum cosmology with cylindrical phase space is recovered in 
the limit  $R_2\rightarrow\infty$. In such a case, the so-called polymerization of the 
momentum $p$, with the polymerization scale $\lambda= \frac{1}{R_1}$, is obtained.  

Furthermore, equations of motion for the minisuperspace model with the spherical phase 
space can be solved analytically. Analysis of the solutions confirmed that both UV and IR results 
are expected due to compactness of the phase space (see also discussion in Ref. \cite{Nozari:2014qja}). The UV 
effects are associated with the $p$ variable (related to the Hubble factor) while the IR effects are associated 
with the $q$ variable (related to the scale factor). In the considered model, manifestation of  
the IR effects is the maximal possible value of $q$ associated with the phase of recollapse 
of the Universe. On the other hand, the UV effects lead to renormalization of the cosmological 
constant and existence of their maximal value $\Lambda_*$. 

Thereafter, we have analyzed a fully quantum version of the theory and promoted the constraint 
to be a quantum operator. It turned out that the constraint can be solved recursively for any 
value of the quantum number $s$.  However, the nontrivial solutions exist only in the bosonic
case, while in the fermionic case the solutions may exist only for some special values of the 
cosmological constant $\Lambda$. As an example we have analyzed the case with $s=2$.
We have found both physical states (one for $\delta\neq 7/18$ and three for $\delta=7/18$) 
that satisfy the constraint as well deriving the form of the projector operator $\hat{P}$. We used 
the projection operator to evaluate transition amplitude between two coherent states that are 
associated with the end points of the classical trajectory. We have found that the probability of 
transition is maximal for $\Lambda=0$.

Some of the future steps in the direction of the research are as follows:
\begin{itemize}
\item {\bf Analysis of the de Sitter model with positive curvature}. This can resolve a problem with the 
interpretation of the IR limit since, in this case, the spatial volume is finite and well defined. 
Furthermore, the model is expected to lead to oscillatory cosmological evolution. 
\item {\bf Analysis of the toroidal compact phase space $\mathbb{S}\times\mathbb{S}$}. 
This case can be viewed as a symmetry-reduced version of the $SU(2)\times SU(2)$ phase space, 
being a possible generalization of the loop quantum gravity phase space. 
\item {\bf Reconstruction of the physical Hamiltonian}. It would be worth reconstructing and studying 
properties of the physical Hamiltonian associated with the constrained system under investigation. 
The resulting physical spin Hamiltonian can be used to implement (in laboratory) analog spin 
models corresponding to the gravitational dynamics under investigation. For this purpose, e.g. 
tunable magnetic metamaterials with controlled spins could be used. Such systems have been 
successfully implemented experimentally in the two-dimensional case \cite{Louis}. 
\item {\bf Taking into account matter}. Our focus on this article was on the gravitational degrees
of freedom and contribution from the cosmological constant has only been taken into account. 
This needs to be generalized by investigating contributions from different 
forms of matter. Especially interesting in the cosmological context is the case of a scalar field. 
Two effects have to be taken into account. The first is an appropriate modification of the 
scalar field Hamiltonian such that the compactness associated with the $q$ variable is introduced in a 
consistent way. Another issue is the compactness of the scalar field itself. This second
issue has be investigated in Ref. \cite{Mielczarek:2017ny}. However, consistent merging of 
both types of compactness (gravitational and scalar field) is an open issue to be addressed
in future studies. 
\item {\bf Semi-classical limit}. In Sec. \ref{QuantumConsiderations} the quantum counterpart 
of the compact phase space minisuperspace model has been introduced and investigated. 
A physical quantum state belonging to the kernel by the constraint can be found. One can 
expect that from the state, the classical cosmological dynamics should be recovered in the 
(semiclassical) large spin limit, $s\rightarrow \infty$. In particular, transition amplitudes between 
two $SU(2)$ coherent states in the  $s\rightarrow \infty$ limit have to be investigated. Furthermore, 
analysis of the transitions amplitudes should be extended beyond the case of boundaries of
a classical trajectory, e.g., to the case of arbitrary two points on the phase space. 
\\
\end{itemize}

\section*{Acknowledgements}

J. M. is supported by the Sonata Bis Grant No. DEC-2017/26/E/ST2/00763 of the National Science 
Centre Poland and the Mobilno\'s\'c Plus Grant No. 1641/MON/V/2017/0 of the Polish Ministry of 
Science and Higher Education.

\end{document}